\documentclass{article}
\usepackage{spconf,amsmath,graphicx}

\title{End-to-end neural speech coding for real-time communications}
\name{Xue Jiang$^{1 \ast}$, Xiulian Peng$^{2}$, Chengyu Zheng$^{1}$, Huaying Xue$^{2}$, Yuan Zhang$^{1}$, Yan Lu$^{2}$}
\address{$^{1}$  Communication University of China, Beijing, China \\
$^{2}$  Microsoft Research Asia, Beijing, China}

\begin{document}
%\ninept
%
\maketitle

\renewcommand{\thefootnote}{\fnsymbol{footnote}}
\footnotetext[1]{This work was done when Xue Jiang was an intern at MSRA.} 
\begin{abstract}
Deep-learning based methods have shown their advantages in audio coding over traditional ones but limited attention has been paid on real-time communications (RTC). This paper proposes the TFNet, an end-to-end neural speech codec with low latency for RTC. It takes an encoder-temporal filtering-decoder paradigm that has seldom been investigated in audio coding. An interleaved structure is proposed for temporal filtering to capture both short-term and long-term temporal dependencies. Furthermore, with end-to-end optimization, the TFNet is jointly optimized with speech enhancement and packet loss concealment, yielding a one-for-all network for three tasks. Both subjective and objective results demonstrate the efficiency of the proposed TFNet.  
\end{abstract}
\section{Introduction}
\label{sec:intro}

Deep learning has shown its great advantages in a wide range of applications including natural language processing, computer vision, etc. In recent years, such gains are observed in neural audio coding as well. Existing neural audio codecs could mainly be grouped into two categories, generative decoder models \cite{wavcodec,Lyra,sampleRNN,generative,improveopus,LPCNet} and end-to-end neural audio coding \cite{VQ-VAE-wavenet,disentangle,soundstream,cascaded}. The former leverages generative models such as WaveNet \cite{wavcodec}, its variants WaveGRU in Lyra \cite{Lyra}, WaveRNN in LPCNet \cite{LPCNet} and SampleRNN \cite{sampleRNN} at the decoder for low-rate speech coding; While for encoding, handcrafted features are extracted from a speech signal, which are typically by a standard parametric codec or a transformed log mel-spectra. These algorithms have clearly outperformed traditional codecs at a very low rate for speech coding. Further, some researchers introduce VQ-VAE \cite{VQ-VAE} into compression and opens the door for end-to-end neural speech coding \cite{VQ-VAE-wavenet,disentangle}. Learned phone embeddings are encoded and a WaveNet is used at the decoder conditioning on the embedding to restore speech signal. These methods typically have a high complexity with auto-regressive networks. Without restricted by speech, some researchers propose to use autoencoders with vector quantization for end-to-end neural audio coding \cite{soundstream,cascaded}. A typical encoder is employed to extract features for encoding with vector quantization and a decoder is utilized for recovering the original audio waveform from quantized features. This category makes a versatile neural audio codec possible; however, its potential is yet far from full exploration. Among all these methods, they mostly still target at coding efficiency with little attention on low latency and error resilience for real-time communications.

\begin{figure}[tb]
\begin{minipage}[b]{1.0\linewidth}
  \centering
  \centerline{\includegraphics[width=7.5cm]{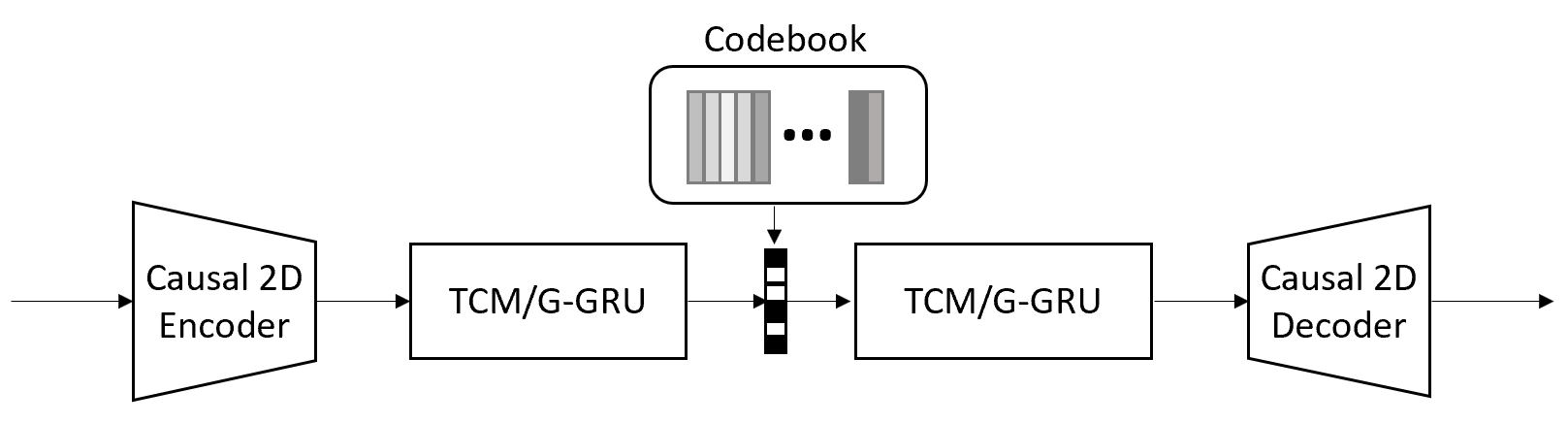}}
%  \vspace{2.0cm}
  \centerline{(a) TFNet neural codec}\medskip
\end{minipage}
\begin{minipage}[b]{1.0\linewidth}
  \centering
  \centerline{\includegraphics[width=9.0cm]{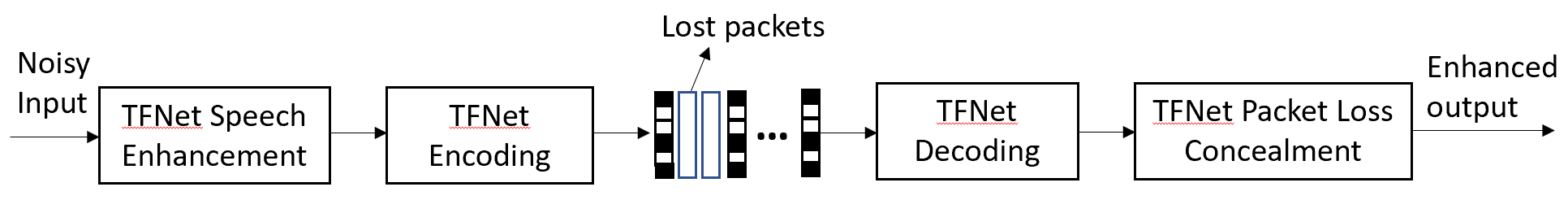}}
  \centerline{(b) Cascaded network for joint optimization}\medskip
\end{minipage}
\hfill
\begin{minipage}[b]{1.0\linewidth}
  \centering
  \centerline{\includegraphics[width=6.5cm]{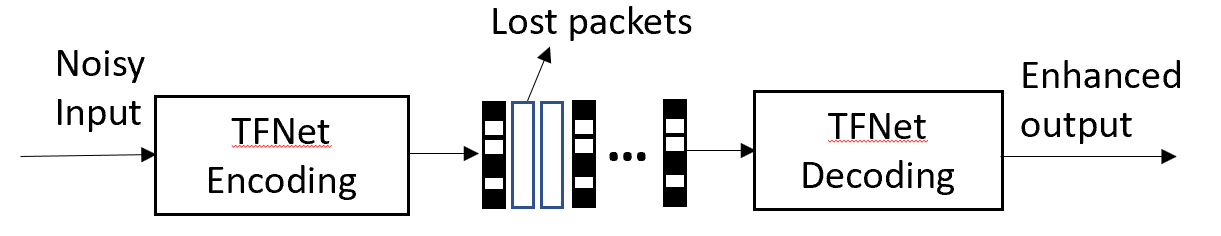}}
  \centerline{(c) All-in-one network for joint optimization}\medskip
  \vspace{-0.2cm}
\end{minipage}
\caption{TFNet neural codec and its joint optimization for real-time communications.}
\vspace{-0.2cm}
\label{fig:tfnet}
\end{figure}

In light of the success of the encoder-temporal filtering-decoder paradigm based on causal convolutions commonly used for regression tasks in real-time audio signal processing \cite{TCNN,CRN}, in this paper we propose the TFNet, a low-latency neural speech codec with end-to-end optimization for real-time communications, following this paradigm (see Fig.\ref{fig:tfnet}(a)). We show that this paradigm works well for low-latency neural speech coding. Particularly, an interleaved structure with causal temporal convolution module (TCM) and group-wise gated recurrent unit (G-GRU) is proposed for temporal filtering for joint long-term and short-term correlation exploitation. Although taking speech as an example in this paper, the TFNet could be easily extended to other audio types such as music as well. 

Furthermore, we exploit the optimization of codec under two types of degradation commonly existing in real-time communications, i.e. background noises and packet losses. We show that with the end-to-end optimization brought by audio coding, user-perceived audio quality could be optimized more efficiently by joint optimization instead of three separate TFNet models optimized for each local target. Particularly, we show that a single TFNet model performs comparatively with three cascaded models under joint optimization (see Fig.\ref{fig:tfnet}). This investigation also shows the efficiency of TFNet for both separation (speech enhancement) and restoration (packet loss concealment and coding) tasks.

%\begin{figure*}[tb]
%\begin{minipage}[b]{1.0\linewidth}
%  \centering
%  \centerline{\includegraphics[width=17.5cm]{figures/modules.PNG}}
%\vspace{-0.4cm}
%\end{minipage}
%\caption{Different causal temporal filtering modules. (a) Dilated temporal convolution module (TCM); (b) Gated recurrent unit (GRU); (c) Causal temporal self-attention module (TSA); (d) Group-wise GRU; (e) Group-wise TSA.}
%\label{fig:temporal}
%\end{figure*}

\begin{figure}[tb]
\begin{minipage}[b]{1.0\linewidth}
  \centering
  \centerline{\includegraphics[width=8.5cm]{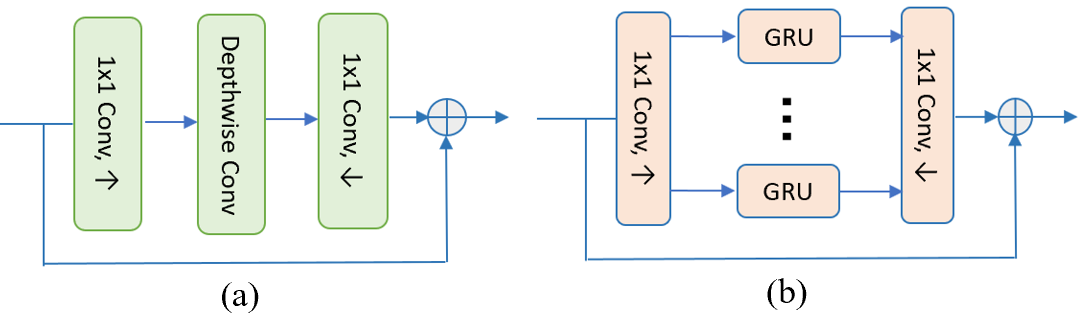}}
%\vspace{-0.4cm}
\end{minipage}
%\vspace{-0.4cm}
\caption{Two causal temporal filtering modules. (a) Dilated temporal convolution module (TCM); (b) Group-wise gated recurrent unit (G-GRU).}
\vspace{-0.4cm}
\label{fig:temporal}
\end{figure}

The rest of the paper is organized as follows. Section 2 expains the details of the proposed TFNet codec. The joint optimization with speech enhancement and packet loss concealment is described in Section 3. Section 4 shows the experimental results and finally Section 5 concludes this paper. 
\vspace{-0.2cm}

%\cite{clone-wavenet}
%\cite{Enhancecodec,handlenoise,soundstream}

\section{Proposed TFNet Codec}
\vspace{-0.2cm}
\subsection{Overview}
The TFNet-based codec takes the time-frequency spectrum (20ms window with a 5ms hop length) as input with a power-law compression on the amplitude before feeding into the network. As the dynamic range of speech is high due to harmonics, the compression performs as a kind of input normalization so that the importances of different frequencies are balanced and the training is more stable. 

The network consists of an encoder and several temporal filtering blocks for encoding (see Fig.\ref{fig:tfnet}(a)). Among them, the encoder exploits local two-dimensional (2D) correlations and the temporal filtering blocks exploit longer-term temporal dependencies with past frames for feature extraction. The two-level feature extraction is important in learning to extract features with good representation capability, error resilience to packet losses and to possibly remove undesired information such as background noises if wanted. The learned features are then quantized through a learned vector quantizer and coded in fixed-length coding. 

For decoding, there are several temporal filtering blocks followed by a decoder for reconstruction. Considering the packet losses in real-time communications, the decoding needs to be resilient to this losses with recovery capability and minimum error propagation. Therefore, a heterogeneous structure is designed with more temporal filtering blocks for decoding than encoding. 

The whole network is end-to-end trained to optimize the reconstruction quality under a rate constraint. All convolutions are causal in temporal dimension so that it can keep a low latency of 20ms.

\vspace{-0.2cm}
\subsection{Encoder and Decoder}
The encoder consists of several causal 2D convolutional layers, each followed by a batch normalization (BN) and a parametric ReLU (PRelu) for nonlinearity. After each convolutional layer, the feature is downsampled by 2 or 4 in frequency dimension and finally all frequency information is folded into channels. Let $X^I\in R^{T\times F\times 2}$ denote the input feature. After the encoder, the feature is $X^E\in R^{T\times 1\times C}$ for feeding into temporal filtering blocks. $T$, $F$ and $C$ are number of frames, frequency bins and channels, respectively. Convolutions are causal along the temporal dimension so $T$ is kept without any downsampling. The decoder is symmetric to the encoder with causal 2D deconvolutional layers. After the decoder, it outputs the reconstructed complex spectrum $X^R\in R^{T\times F\times 2}$ with an inverse STFT to get the output waveform.

\vspace{-0.2cm}
\subsection{Temporal Filtering}
We utilize two types of temporal filtering blocks as shown in Fig.\ref{fig:temporal}, the dilated temporal convolution module (TCM) and the group-wise gated recurrent unit (G-GRU). Both are causal and low-complexity. The TCM module as that in \cite{conv-tasnet} consists of two convolutions with a kernel size of $1\times 1$ to change channel dimensions and dilated depthwise convolutions to exploit temporal correlations with low complexity. Several TCM blocks with different dilation rates are grouped as a large block to increase the receptive field and diversities. The group-wise GRU blocks in Fig.\ref{fig:temporal} (b) split channels into $N$ groups and leverage temporal dependencies inside each group independently. This group-wise GRU variants not only reduce the complexity but also increase the flexibility and representation capability for providing frequency-aware temporal filtering as channels are learned from frequencies. 

As TCM is more efficient in exploring short-term and middle-term temporal evolutions while GRU is mostly proposed to capture long-term dependencies, we propose to combine them in an interleaved way for capturing short-term and long-term temporal correlations at different depths. Our experimental results verify that the interleaved structures are more efficient than a single one, which will be presented in Section 4.

%Heterogeous design (to explore what kind of temporal filtering used for encoding and decoding, the choices would be different from that for codec only)

\vspace{-0.2cm}
\subsection{Vector Quantization}
The vector quantizer discretizes the learned features in encoding with a set of learnable codebooks according to the target bitrate. Before quantization, the features after encoding $X^S\in R^{T\times 1\times C}$ are reduced to $X^{Q}\in R^{T\times 1\times C'}$ through a $1\times 1$ convolution ($C' < C$). We take a group quantization by splitting channels $C'$ into $N$ groups and coding each group by an independent codebook. Let $S$ denote the number of codewords in each codebook and $K=C'/N$ the dimension of each codeword. In the proposed scheme, a window length of 20ms and hop length of 5ms is adopted for STFT and thus the bitrate is given by $N\times log_2 S /5$ kbps. For 6kbps, $C'$, $N$, $S$ and $K$ are set to 120, 3, 1024, and 40, respectively. The codebooks are learned with exponential moving average, following that in \cite{VQ-VAE}. The quantized features $\hat{X}^{Q}\in R^{T\times 1\times C'}$ are enlarged to the shape $T\times 1\times C$ before feeding into the temporal filtering blocks in decoding.

\vspace{-0.2cm}
\subsection{Loss Function}
The loss function is a combination of two terms $\mathcal{L}=\mathcal{L}_{recon}+\alpha \mathcal{L}_{VQ}$. $\mathcal{L}_{recon}$ is the reconstruction loss and the other $\mathcal{L}_{VQ}$ puts a constraint on vector quantization. We use a mean-square error on the power-law compressed spectrum between the original and the decoded signals for reconstruction loss \cite{cocktail}. To ensure STFT consistency \cite{stft-consistency}, the decoded spectrum is first transformed into waveform domain through an inverse STFT and then transformed into TF-domain again through a STFT to calculate the loss. The second term $\mathcal{L}_{VQ}$ is the commitment loss used in VQ-VAE, which forces the encoder to generate a representation close to its codeword. $\alpha$ is a weighting factor to balance the two terms. 

\section{Joint Optimization with Speech Enhancement and Packet Loss Concealment}
In real-time communications, there are several types of degradations besides quality loss by audio coding, such as background noises and packet losses. Owing to the end-to-end learnable audio codec, it is feasible to jointly optimize the audio coding with speech enhancement (SE) and packet loss concealment (PLC). We consider two ways of joint optimization, the cascaded network with an enhancer before the codec and a PLC network after it (see Fig.\ref{fig:tfnet}(b)), and the all-in-one network that takes the similar network structure as the codec but optimized for noisy input with packet losses (see Fig.\ref{fig:tfnet}(c)).
%Along this line, Casebeer et al. propose the joint optimization of audio coding with speech enhancement \cite{Enhancecodec} but it is only compared with a light-weight enhancer.  

\vspace{-0.2cm}
\subsection{Cascaded Network}
The cascaded network consists of three modules, an enhancer for pre-processing, an audio codec and a PLC network for post-processing. As speech is more efficient in compression than a noisy audio, the enhancer is put before the codec. The three modules are all based on TFNet-like structures and jointly trained in an end-to-end way.

The pre-processing enhancer takes noisy audio as input and outputs enhanced audio for feeding into the codec. Different from the TFNet-based codec, there are skip connections between the encoder and the decoder in the enhancer to get rid of information loss. We employ causal gated blocks in decoder and output an amplitude gain and the phase for reconstruction, similar to that in \cite{sn-net}.

Under packet losses, the neural codec is adjusted in that in decoding it takes both the quantized features with lost packets as zero and a mask showing where the loss happens as input. The mask is also injected into each temporal filtering blocks in decoding. The post-processing PLC module operates in the waveform domain, taking a TFNet-based structure with both the decoded audio and the mask as input. There are also skip connections in the PLC network as that in enhancer. As a restoration task, it outputs a complex residue in TF domain which is added into the spectrum of the decoded audio for reconstruction.

For training, the three networks are concatenated and jointly trained from end to end. For better quality, we take a two-stage training. First, the enhancer and the codec are separately trained with noisy and clean data, respectively. Then the cascaded network is finetuned from that, with two additional supervisions at the output of the enhancer and the codec, respectively, using the same reconstruction loss as $\mathcal{L}_{recon}$.

\vspace{-0.2cm}
\subsection{All-in-One Network}
The all-in-one network is resilient to both background noises and packet losses with only a single codec network. To accommodate packet losses, the decoding part in the codec is adjusted similar to that in the cascaded network. It is trained from scratch with an auxiliary supervision added for the encoding part to remove noises for efficient coding. This is achieved by adding a decoder after the temporal filtering blocks of the encoding, which is forced to output clean audio in training. During inference, this decoder is not needed.

\section{Experimental Results}
\vspace{-0.2cm}
\subsection{Datasets, Settings and Evaluation Metrics}
We synthesized 890 hours of 16khz noisy audios with clean speech, noises and room impulses from the Deep Noise Suppression Challenge at ICASSP 2021 \cite{DNSchallenge}. The clean audio includes multilingual speech, emotional and singing clips. The SNR is randomly chosen inbetween -5dB and 20dB and the speech level within -40 to -10dB. Each audio is cut into 3-second segments for training. The speech enhancement performs both denoising and dereverberation. The packet losses are simulated following the three-state model \cite{threestate}. For testing, we take 1400 audios each with 10 seconds long without any overlapping with training data.

During training, we use the adam optimizer with a learning rate of 0.0004. The network is trained for 100 epochs with a batch size of 200. 

In evaluation, except the subjective listening test in Section 4.2, three metrics are used for ablation studies in Section 4.3 and 4.4, the wideband PESQ \cite{pesq}, STOI \cite{stoi} and the DNSMOS \cite{DNSMOS}. Although they are not designed and optimized for exactly the same task, we found that for the same kind of distortions in all compared schemes, they match well with perceptual quality.

\begin{figure}[tb]
\centering
\centerline{\includegraphics[width=7.5cm]{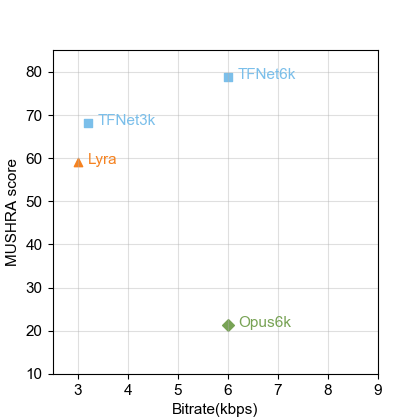}}
\vspace{-0.2cm}
\caption{Subjective evaluation results (Reference score: 88).}
\vspace{-0.2cm}
\label{fig:subjective}
\end{figure}

\vspace{-0.2cm}
\subsection{Comparison with other Codecs}
The codec network is trained and measured on the clean data only in this section. We conduct a subjective listening test with a MUSHRA-inspired crowd-sourced method \cite{subjective}. There are 10 participants, each evaluates 12 samples. We compare the TFNet-based neural codec with Lyra\footnote{https://github.com/google/lyra} and Opus\footnote{https://opus-codec.org}, two codecs used for real-time communications. The former is a neural speech codec proposed in this year. As shown in Fig.\ref{fig:subjective}, the TFNet at around 3kbps clearly outperforms Lyra at 3kbps and TFNet at 6kbps is much better than Opus at 6kbps, which demonstrates its superiority.

\vspace{-0.2cm}
\subsection{Evaluation on Joint Optimizations}

We evaluate the joint optimization of codec, speech enhancement and PLC in this section using the noisy/clean paired data with simulated packet loss traces. Three methods are compared here, a baseline with separately trained enhancement, coding and PLC models, the cascaded network and the all-in-one network. In baseline, coding and PLC networks are trained on raw clean data only. The enhancer and PLC networks have 470K parameters and 1.2M MACs per 20ms, far less than the codec network with 5M parameters. 

In Table \ref{tab:jointdns} and \ref{tab:joint}, the comparison results on two and three-task joint optimizations are presented, respectively. It is observed that the two joint optimization methods clearly outperforms the baseline in all metrics. Although no pre-processing nor post-processing networks are used, the all-in-one network performs competitively with the cascaded one, showing the strong discrimination and representation capability of TFNet. Another observation is that the PLC network trained on raw clean data in baseline method is sensitive to mismatch in the input.

\vspace{-0.2cm}
\subsection{Evaluation on Temporal Filtering Modules}
We compare the interleaved structure in TFNet neural codec with two modules TCM and GRU commonly used in regression tasks of speech enhancement here. All schemes are compared under the same computational complexity with 1.4M parameters and 3.3M MACs for each 20ms window for encoding and decoding. In this study, all temporal filtering modules are used for decoding only to evaluate their recovery capability. Table \ref{tab:temporal} shows the comparison results. It can be seen that the interleaved structure performs the best for capturing both short-term and long-term temporal correlations. 

\begin{table}[t]
  \centering
  \fontsize{8}{10}\selectfont
  \setlength\abovedisplayskip{1pt}
  \setlength\belowdisplayskip{1pt}
  \label{tab:jointdns}
  \caption{Comparison of coding efficiency under background noises.}
  \vspace{0.2cm}
\begin{tabular}{|c|c|c|c|c|c|}
\hline
 \  & PESQ  &STOI & DNSMOS \\  
%\multicolumn{1}{|c|}{ $\lambda_1$} &  \multicolumn{1}{c|}{0}&  \multicolumn{1}{c|}{0.01} &  \multicolumn{1}{c|}{0.1} &  \multicolumn{1}{c|}{0.5}&\multicolumn{1}{c|}{1} &  \multicolumn{1}{c|}{2}& \multicolumn{1}{c|}{10}   \\ 
\hline
 Baseline &1.740 & 0.811 & 3.29  \\  
 \hline
 Cascaded &1.777 & \bf{0.822} & \bf{3.51} \\  
 \hline
 All-in-one &\bf{1.794} & 0.821 & 3.48 \\  
 \hline
\end{tabular}
\vspace{-0.2cm}
\label{tab:jointdns}
\end {table}

\begin{table}[t]
  \centering
  \fontsize{8}{10}\selectfont
  \setlength\abovedisplayskip{1pt}
  \setlength\belowdisplayskip{1pt}
  \label{tab:joint}
  \caption{Comparison of coding efficiency under background noises and packet losses.}
  \vspace{0.2cm}
\begin{tabular}{|c|c|c|c|c|c|}
\hline
 \  & PESQ  &STOI & DNSMOS \\  
%\multicolumn{1}{|c|}{ $\lambda_1$} &  \multicolumn{1}{c|}{0}&  \multicolumn{1}{c|}{0.01} &  \multicolumn{1}{c|}{0.1} &  \multicolumn{1}{c|}{0.5}&\multicolumn{1}{c|}{1} &  \multicolumn{1}{c|}{2}& \multicolumn{1}{c|}{10}   \\ 
\hline
 Baseline &1.413 & 0.747 & 3.04  \\  
 \hline
 Cascaded &\bf{1.545} & \bf{0.778} & \bf{3.38} \\  
 \hline
 All-in-one &1.510 & 0.770 & 3.33 \\ 
 \hline
\end{tabular}
\vspace{-0.2cm}
\label{tab:joint}
\end {table}

\vspace{-0.2cm}
\begin{table}[ht]
  \centering
  \fontsize{8}{10}\selectfont
  \setlength\abovedisplayskip{1pt}
  \setlength\belowdisplayskip{1pt}
  \label{tab:temporal}
  \caption{Comparison of different temporal filtering modules.}
  \vspace{0.2cm}
\begin{tabular}{|c|c|c|}
\hline
 \  & PESQ  &STOI \\  
%\multicolumn{1}{|c|}{ $\lambda_1$} &  \multicolumn{1}{c|}{0}&  \multicolumn{1}{c|}{0.01} &  \multicolumn{1}{c|}{0.1} &  \multicolumn{1}{c|}{0.5}&\multicolumn{1}{c|}{1} &  \multicolumn{1}{c|}{2}& \multicolumn{1}{c|}{10}   \\ 
\hline
 TCM & 2.447 &0.869   \\  
 \hline
 GRU & 2.360 &0.863 \\  
 \hline
 Interleave-TCM-G-GRU &\bf{2.501} &\bf{0.870} \\  
 \hline
\end{tabular}
\vspace{-0.4cm}
\label{tab:temporal}
\end {table}

%\begin{table}[ht]
%  \centering
%  \fontsize{8}{10}\selectfont
%  \setlength\abovedisplayskip{1pt}
%\setlength\belowdisplayskip{1pt}
%  \label{tab:temporal}
%      \caption{Comparison of different temporal filtering modules.}
%\begin{tabular}{|c|c|c|c|c|c|}
%\hline
% \  & PESQ  &STOI & DNSMOS \\  
%\multicolumn{1}{|c|}{ $\lambda_1$} &  \multicolumn{1}{c|}{0}&  \multicolumn{1}{c|}{0.01} &  \multicolumn{1}{c|}{0.1} &  \multicolumn{1}{c|}{0.5}&\multicolumn{1}{c|}{1} &  \multicolumn{1}{c|}{2}& \multicolumn{1}{c|}{10}   \\ 
%\hline
% TCM &3.35 & 2.94 & 2.59  \\  
% \hline
% GRU &3.96 & 3.72 & 3.36 \\  
% \hline
% Interleave-TCM-GRU &3.91 &3.74 & 3.34 \\  
% \hline
% Interleave-TCM-TSA &4.06 &3.82 &3.48 \\  
% \hline
% Interleave-TCM-groupGRU &\bf{4.14} &\bf{3.83} &\bf{3.54} \\  
% \hline
% Interleave-TCM-groupTSA &4.06 &3.82 &3.48 \\
% \hline
%\end{tabular}
%\label{tab:temporal}
%\end {table}

\section{Conclusions}
We propose the TFNet, a low-latency neural speech codec with end-to-end optimization. Taking it as a backbone, we further investigate the joint optimization of the codec with speech enhancement and packet loss concealment. Experimental results demonstrate the representation capability of TFNet in speech coding and regression tasks. As the network itself makes no assumption on the nature of the audio signal it encodes and no explicit speech-specific features are extracted, it can be extended for other audio types as well in the future.

%\vfill\pagebreak

\bibliographystyle{IEEEbib}
\bibliography{main}

\begin{thebibliography}{10}

\bibitem{wavcodec}
W.B. Kleijin, F.S. Lim, A.~Luebs, and J.~Skoglund,
\newblock ``Wave{N}et based low rate speech coding,''
\newblock in {\em ICASSP}. IEEE, 2018, pp. 676--680.

\bibitem{Lyra}
W.B. Kleijn, A.~Storus, M.~Chinen, T.~Denton, F.S.C. Lim, A.~Luebs,
  J.~Skoglund, and H.~Yeh,
\newblock ``Generative speech coding with predictive variance regularization,''
\newblock in {\em arXiv:2102.09660}, 2021.

\bibitem{sampleRNN}
J.~Klejsa, P.~Hedelin, C.~Zhou, R.~Fejgin, and L.~Villemoes,
\newblock ``High-quality speech coding with sample {RNN},''
\newblock in {\em ICASSP}. IEEE, 2019, pp. 7155--7159.

\bibitem{generative}
R.~Fejgin, J.~Klejsa, L.~Villemoes, and C.~Zhou,
\newblock ``Source coding of audio signals with a generative model,''
\newblock in {\em ICASSP}. IEEE, 2020, pp. 341--345.

\bibitem{improveopus}
J.~Skoglund and J.M. Valin,
\newblock ``Improving {O}pus low bit rate quality with neural speech
  synthesis,''
\newblock in {\em Interspeech}, 2020.

\bibitem{LPCNet}
Valin J.M. and Skoglund J.,
\newblock ``{LPCNet}: improving neural speech synthesis through linear
  prediction,''
\newblock in {\em ICASSP}. IEEE, 2019.

\bibitem{VQ-VAE-wavenet}
C.~G\^{a}rbacea, A.~van~den Oord, Y.~Li, F.S. Lim, A.~Luebs, O.~Vinyals, and
  T.~C. Walters,
\newblock ``Low bit-rate speech coding with {VQ-VAE} and a {W}ave{N}et
  decoder,''
\newblock in {\em 2019 IEEE Int. Conf. Acoust Speech Signal Processing
  (ICASSP)}. IEEE, 2019, pp. 735--739.

\bibitem{disentangle}
J.~Williams, Yi~Zhao, E.~Cooper, and J.~Yamagishi,
\newblock ``Learning disentangled phone and speaker representations in a
  semi-supervised {VQ-VAE} paradigm,''
\newblock in {\em 2021 IEEE Int. Conf. Acoust Speech Signal Processing
  (ICASSP)}. IEEE, 2021.

\bibitem{soundstream}
N.~Zeghidour, A.~Luebs, A.~Omran, J.~Skoglund, and M.~Tagliasacchi,
\newblock ``Sound{S}tream: an end-to-end neural audio codec,''
\newblock in {\em arXiv:2107.03312v1}, 2021.

\bibitem{cascaded}
K.~Zhen, J.~Sung, M.S. Lee, S.~Beack, and M.~Kim,
\newblock ``Cascaded cross-module residual learning towards lightweight
  end-to-end speech coding,''
\newblock in {\em Proceedings of the Annual Conference of the International
  Speech and Communication Association (Interspeech)}, 2019.

\bibitem{VQ-VAE}
Aaron van~den Oord, Oriol Vinyals, and Koray Kavukcuoglu,
\newblock ``Neural discrete representation learning,''
\newblock {\em arXiv preprint arXiv:1711.00937}, 2017.

\bibitem{TCNN}
A.~Pandey and D.L. Wang,
\newblock ``T{CNN}: Temporal convolutional neural network for real-time speech
  enhancement in the time domain,''
\newblock in {\em 2019 IEEE Int. Conf. Acoust Speech Signal Processing
  (ICASSP)}. IEEE, 2019.

\bibitem{CRN}
K.~Tan and D.L. Wang,
\newblock ``A convolutional recurrent neural network for real-time speech
  enhancement,''
\newblock in {\em Proceedings of the Annual Conference of the International
  Speech and Communication Association (Interspeech)}, 2018.

\bibitem{conv-tasnet}
Yi~Luo and Nima Mesgarani,
\newblock ``Conv-tasnet: Surpassing ideal time--frequency magnitude masking for
  speech separation,''
\newblock {\em IEEE/ACM transactions on audio, speech, and language
  processing}, vol. 27, no. 8, pp. 1256--1266, 2019.

\bibitem{cocktail}
Ariel Ephrat, Inbar Mosseri, Oran Lang, Tali Dekel, Kevin Wilson, Avinatan
  Hassidim, William~T Freeman, and Michael Rubinstein,
\newblock ``Looking to listen at the cocktail party: A speaker-independent
  audio-visual model for speech separation,''
\newblock {\em arXiv preprint arXiv:1804.03619}, 2018.

\bibitem{stft-consistency}
Scott Wisdom, John~R Hershey, Kevin Wilson, Jeremy Thorpe, Michael Chinen,
  Brian Patton, and Rif~A Saurous,
\newblock ``Differentiable consistency constraints for improved deep speech
  enhancement,''
\newblock in {\em ICASSP}. IEEE, 2019, pp. 900--904.

\bibitem{sn-net}
C.~Zheng, X.~Peng, Y.~Zhang, S.~Srinivasan, and Y.~Lu,
\newblock ``Interactive speech and noise modeling for speech enhancement,''
\newblock in {\em AAAI}, 2021.

\bibitem{DNSchallenge}
C.~K~A Reddy, H.~Dubey, V.~Gopal, R.~Cutler, S.~Braun, H.~Gamper, R.~Aichner,
  and S.~Srinivasan,
\newblock ``I{CASSP} 2021 deep noise suppression challenge,''
\newblock {\em arXiv preprint arXiv:2009.06122}, 2020.

\bibitem{threestate}
B.P. Milner and A.B. James,
\newblock ``An analysis of packet loss models for distributed speech
  recognition,''
\newblock {\em Proceedings INTERSPEECH, 8th Internaltional Conference on Spoken
  Language Processing}, 2004.

\bibitem{pesq}
ITUT Rec,
\newblock ``P.862.2: Wideband extension to recommendation p.862 for the
  assessment of wideband telephone networks and speech codecs,''
\newblock {\em International Telecommunication Union, CH--Geneva}, 2005.

\bibitem{stoi}
C.H.Taal, R.C.Hendriks, R.Heusdens, and J.Jensen,
\newblock ``A short-time objective intelligibility measure for time-frequency
  weighted noisy speech,''
\newblock in {\em ICASSP}, 2010.

\bibitem{DNSMOS}
C.~K~A Reddy, V.~Gopal, and R.~Cutler,
\newblock ``D{NSMOS}: a non-intrusive perceptual objective speech quality
  metric to evaluate noise suppressors,''
\newblock in {\em 2021 IEEE Int. Conf. Acoust Speech Signal Processing
  (ICASSP)}. IEEE, 2021.

\bibitem{subjective}
ITU-R,
\newblock ``Recommendation {BS}.1534-1: {M}ethod for the subjective assessment
  of intermediate quality level of audio systems,''
\newblock {\em International Telecommunication Union Radiocommunication
  Assembly}, 2001.

\end{thebibliography}

\end{document}